\documentstyle[12pt]{article}

\newcommand{\be}{\begin{equation}}
\newcommand{\ee}{\end{equation}}
\newcommand{\bes}{\begin{eqnarray}}
\newcommand{\ees}{\end{eqnarray}}
\newcommand{\bea}{\begin{equation}\begin{array}{lcl}}
\newcommand{\eea}[1]{\end{array}\label{#1}\end{equation}}
\newcommand{\ba}{\begin{array}}
\newcommand{\ea}{\end{array}}

\newcommand{\refe}[1]{(\ref{#1})}
\newcommand{\ns}{\normalsize}
\newcommand{\ler}{\stackrel{\scriptstyle <}{\scriptstyle\sim}}
\newcommand{\la}{\lambda}
\newcommand{\La}{\Lambda}
\newcommand{\De}{\Delta}
\newcommand{\al}{\alpha}
\newcommand{\bt}{\beta}
\newcommand{\ga}{\gamma}
\newcommand{\nn}{\nonumber}
\newcommand{\ra}{\rightarrow}
\newcommand{\ep}{\epsilon}
\newcommand{\de}{\delta}
\newcommand{\lra}{\leftrightarrow}
\begin{document}
\begin{titlepage}
\setcounter{page}{1}
\title{{\large \bf NJL breaking of supersymmetric GUTs}
         \thanks{supported by Deutsche Forschungsgemeinschaft}}
\author{{\ns E.~J.~Chun
         \thanks{supported by a KOSEF fellowship and the CEC science
                                        project no.~SC1--CT91--0729.}
        \hspace{0.02cm} and
        A.~Lukas}\\[0.2cm]
        {\ns Physik Department}\\
        {\ns Technische Universit\"at M\"unchen}\\
        {\ns D-8046 Garching, Germany}\\
        {\ns and}\\[0.1cm]
        {\ns Max-Planck-Institut f\"ur Physik}\\
        {\ns Werner-Heisenberg-Institut}\\
        {\ns P.~O.~Box 40 12 12, Munich, Germany}}

\date{{\ns February 1993}}

\maketitle

\setlength{\unitlength}{1cm}
\begin{picture}(5,1)(-12.0,-14)
\put(0,-0.5){TUM - TH - 154/93}
\put(0,-1.0){MPI - Ph/93 - 9}
\put(0,-1.5){hep-ph/9303245}
\end{picture}

\begin{abstract}
We analyze the breakdown of SUSY GUTs driven by Nambu--Jona--Lasinio
condensates. Starting with the most general gauge invariant and pure
K\"ahlerian Lagrangian up to quartic order we solve the one loop gap
equation and determine the breaking direction. This is done for
various classes of groups and spectra of fundamental particles
which especially cover the most promising unifying groups $SU(5)$,
$SO(10)$ and $E_6$. Heavy masses for the fundamental as well as for the
composite particles are calculated. The results are used to single
out candidates which may lead to an acceptable low energy theory.
In these models we discuss some phenomenological aspects and point
out the difficulties in constructing phenomenological viable theories
in our scenario.
\end{abstract}

\thispagestyle{empty}
\end{titlepage}
\clearpage

\setcounter{page}{1}

\section{Introduction}
The unification of gauge coupling constants in the minimal supersymmetric
Standard Model (MSSM)~\cite{amaldi_lang} has directed great attention
on supersymmetric theories. At present Grand--Unified--Theories (GUTs)
represent the only framework able to explain the meeting of coupling
constants. Though GUTs are highly predictive in the gauge
sector large arbitrariness enters the theory through the superpotential
needed for spontaneous breakdown. Especially the explanation of fermion
masses seems to be difficult without adding new structure like discrete
symmetries or ``Ans\"atze'' for the mass matrices to the theory.

Dynamical symmetry breaking (DSB) can improve this situation and in
principle increase the predictivity of GUTs. The Nambu--Jona--Lasinio (NJL)
mechanism~\cite{njl} leads to scalar condensate formation through
an effective four--fermion interaction which can trigger the symmetry
breakdown. It was first shown by Bardeen, Hill and Lindner~\cite{bhl}
that the standard model (SM) can be broken by this mechanism leading
to a prediction
for the top quark mass. However, a fine tuning caused by the quadratic
divergence in the cutoff has to be made in order to obtain small SM masses.
This can be avoided in the supersymmetric generalization of the NJL
mechanism~\cite{buch_ell}. To form condensates supersymmetry
(SUSY) has to be
broken softly and the ``four--fermion--coupling'' has to exceed a
certain critical value. The quadratic dependence on the cutoff $\Lambda$
is then replaced by that of the SUSY breaking scale $\Delta$. Unfortunately,
to reach the critical value the Fermi scale $f$ has to be of the same
order as $\Delta$ which seems to be an unpleasant feature for large
cutoff values. This situation already appears in the dynamically broken
MSSM~\cite{bcl} where a large cutoff is needed to produce acceptable small
top masses. A solution to this problem in the framework of scaleless
$\sigma$--models was proposed by Ellwanger~\cite{ell}. In these models
the scale $f$ is replaced by a dilaton field and therefore determined
dynamically.

Our motivation is to break the GUT symmetry using the NJL mechanism
and to determine the phenomenological properties of the resulting
highly predictive low energy theory.
In this paper, however, we will concentrate on the first step in this
scenario : breaking the GUT symmetry and analyzing basic structures
of the low energy theory. This amounts to minimize the self consistent
gap potential for various classes of groups similar to the analysis
done by Li~\cite{li} for conventional GUTs with an explicit potential.
The difficulties in building models
with phenomenological acceptable low energy limit will already
show up in this discussion. We start with the most general gauge
invariant Lagrangian up to quartic order which is pure K\"ahlerian.
As an effective field theory it can be possibly induced by coset models
or superstring compactifications below Planck scale. Apart from some
connections to coset models we will not specify the origin.
Likewise we do not address how the small Fermi scale can be
produced but just put it as a parameter. In this respect our scenario
differs from a similar one discussed by Ellwanger~\cite{ell} in the context
of an $E_8/SO(10)\times SU(3)\times U(1)$ scaleless coset model.
We adopt the idea to achieve a radiative low energy breaking triggered by
condensate fields which was first proposed in ref.~\cite{buch_ell_1}.

The plan of the paper is as follows. In section 2 the general structures
and formulas are presented. Section 3 is mainly addressed to a discussion
of adjoint condensates as used for the standard GUT breaking. We give
a complete analysis for adjoint condensates formed out of the fundamental
representations of $SU(N)$ and $SO(N)$ and some examples for the second
rank tensors of $SU(N)$. The formation of complex representations
especially for the groups $SO(10)$ and $E_6$ will be the subject of
section 4. In section 5 we give the results for the heavy masses
of the composites and discuss some phenomenological aspects.

\section{General Framework}

Starting point of our analysis is the Langrangian
\be
 {\cal L} = \int d^2 \theta\, d^2 \bar{\theta} \left[
  \frac{1}{n}(\bar{\phi} \phi + \bar{\phi}' \phi ')
  (1- \De^2 \theta^2 \bar{\theta}^2)
  +\frac{c_r^2}{(n f)^2} \bar{\phi}' \la^{A(r)*} \bar{\phi} \phi'
   \la_{A}^{(r)} \phi \right]  \label{start_lag}
\ee
\[
 n = \left\{ \ba{lll} 1&{\rm for}&\phi \ne \phi '\\
                     2&{\rm for}&\phi = \phi ' \ea \right.
\]
where the gauging exponentials have been omitted. The two fields $\phi$ and
$\phi '$ transform as representations ${\bf r}$ and ${\bf r'}$
under the chosen gauge group $G$. Depending on the particle content and
the mass structure of the theory they will be interpreted as new exotic
fields or standard particles. In the latter case we will consider them
as belonging to the heaviest family in the SM. Soft breaking of
supersymmetry is parameterized by the scalar mass $\De^2$.
The group coefficients $\la$
split the interaction term into its irreducible parts (labeled by $r$)
corresponding to the decomposition of ${\bf r}\times {\bf r'}$. We
choose the normalization tr$(\la^+ \la)=n$.
A cutoff $\La$ has to be introduced in order
to define the theory. It refers to new physics at a higher energy scale
which also represents the origin of the relative coupling strengths $c_r^2$.
E.~g.~in coset models related to the breakdown of certain larger groups $G'$
to $G$ the Lagrangian~\refe{start_lag} can be identified with the leading
terms of the K\"ahler potential on the coset space $G'/G$~\cite{coset}.
In such models the ratio of the couplings $c_r$ is determined by
group coefficients.
In order to break our symmetry at the unification scale $M_{\rm GUT} \sim
10^{16}$ GeV we have to demand $\La \gg M_{\rm GUT}$. A ``natural'' choice
is $\La = O(M_{\rm Planck})$. We see that the theory is specified by
a small numbers of parameters, namely $\De^2$, $\La$, $c_r^2$
and the gauge coupling $g$.

To
study
condensate
formation one may introduce two auxiliary fields $H^{(r)A}$ and
$H_{A}^{'(r)}$ in terms of which eq.~\refe{start_lag}
can be rewritten as~\cite{buch_ell}
\bes
 {\cal L} &=& \int d^2 \theta\, d^2 \bar{\theta} \left[
  (\bar{\phi} \phi + \bar{\phi}' \phi ')(1- \De^2 \theta^2 \bar{\theta}^2)
  + \bar{H}^{'(r)A} H_A^{'(r)} \right] \nn \\
  && +\left[ \int d^2 \theta H^{(r)A}(H_A^{'(r)} - \frac{c_r}{n f} \phi'
   \la_{A}^{(r)} \phi)\; +\; {\rm h.\,c.} \right] . \label{aux_lag}
\ees
Eliminating $H$ and $H'$ by their equations of motion leads back to the
Lagrangian~\refe{start_lag} which implies the equivalence of the two
formulations. The missing kinetic term for $H$
can be developed at one--loop--level. This can establish
$H$ and $H'$ as effectively independent degrees of freedom at energies
below the cutoff $\La$. Since only the fields $H$ receive quantum
corrections symmetry breaking is determined by their effective potential. We
define the mass matrix $M$ of the fundamental fields~:
\bes
 m^{(r)A} &=& \frac{1}{f} <H^{(r)A}> \nn \\
 M &=& c_r m^{(r)A} \la_A^{(r)} . \label{m_def}
\ees
In the following we will work in an one--loop approximation for chiral
loop particles taking tree--level gauge effects (i.~e.~D--term effects)
into account. The effective action can then be written as
\be
 V_{\rm eff} = V_{\rm gap} + V_{\rm D} \label{v_eff}
\ee
where $V_{\rm D}$ is the usual D--term
\[
 V_{\rm D} = \frac{1}{2} g^2 D_a D_a, \quad \quad
 D_a = \sum_{r} \bar{H}^{(r)} T_a^{(r)}H^{(r)} .
\]
The NJL self--consistency condition at one loop level leads to the
gap equation
\begin{eqnarray*}
 && \frac{\partial V_{\rm gap}}{\partial m_B^{(r)*}} =  \\
 && f^2 \left[ m^{(r)B} - \frac{c_r}{8 \pi^2 n f^2} {\rm tr}\left(
   \la^{B*(r)} M \left(
  (|M|^2 + \De^2) \ln \frac{\La^2}{|M|^2 + \De^2} - |M|^2
  \ln \frac{\La^2}{|M|^2} \right) \right) \right] = 0 .
  \label{gap_eq}
\end{eqnarray*}
For $\La \gg |M|,\De$ it can be approximated to
\be
 \frac{\partial V_{\rm gap}}{\partial m_B^{(r)*}} =
  f^2 \left[ m^{(r)B} - \frac{c_r\alpha}{n} {\rm tr} \left( \la^{B(r)*} M
  \left(\ln \frac{\La^2}{|M|^2 + \De^2}-1 \right) \right) \right] = 0
  \label{gap_appr}
\ee
with
\[
 \alpha = \frac{\De^2}{8 \pi^2 f^2} .
\]
Neglecting the group structure of eq.~\refe{gap_appr} it can be seen that
a nontrivial solution presumes a small Fermi--scale
$f \ler O(\De)$~\cite{buch_ell}. This expresses the fact that supersymmetry
protects the theory against mass generation. A consequence is that possible
higher dimensional operators are suppressed much weaker than in the
non supersymmetric case. E.~g.~K\"ahler potentials derived from coset models
usually contain higher than quartic terms. Especially VEVs of the order
$M_{\rm GUT}$
- naively inserted in the higher dimensional terms - lead to a disastrous
increasing. We do not see any satisfactory solution to this but we will
just assume that either higher dimensional operators are absent or
suppressed by some large scale $O(\La)$.
For the explicit solution of the gap equation we set $\alpha = 1$ so that
\be
 f^2 = \frac{\De^2}{8 \pi^2} . \label{f_value}
\ee
The interaction strength is then determined by the dimensionless couplings
$c_r = O(1)$. The rough behavior of the VEVs $m$ is given by
\be
 \frac{m^2}{\La^2} \sim \exp (-1/c^2) \label{vev_beh}
\ee
showing that even small changes in $c_r$ (e.~g.~caused by group theoretical
factors) can lead to large differences in the mass scale $m$. To achieve
a breakdown scale save below the cutoff $c\ler 1/2$ has to be chosen.
For a reasonable GUT--scale the value of $c$ should not be much lower.
Integrating the LHS of eq.~\refe{gap_appr} leads to the potential
\be
 V_{\rm gap} =  f^2 \left[ |m|^2 - \frac{1}{n}{\rm tr}
                \left( |M|^2 \ln \frac{\La^2}
                {|M|^2 + \De^2} \right) \right] . \label{pot_appr}
\ee
The minimum value can be obtained by combining eqs.~\refe{gap_appr} and
\refe{pot_appr}~:
\be
 V_{\rm gap,min} = -\frac{f^2}{n} {\rm tr} |M|^2 . \label{pot_min}
\ee

It is well known that in the picture of renormalization compositeness
shows up as a Landau singularity in the Yukawa coupling. One may ask
if three orders of magnitude below the cutoff (assuming $\La\sim
M_{\rm Planck}$ and $M_{\rm GUT}\sim 10^{16}$ GeV) we are already in
the nonperturbative region of this coupling. However, applying the
naive bound $\al_{\rm Yuk}\ler 1$ and running down $\al_{\rm Yuk}$
shows that we are safe below this value for typical examples. On the
other hand we can be sure that we are dealing with a relatively
large coupling $\al_{\rm Yuk}$ and a gauge coupling of
$g^2/4\pi=O(1/25)$. For that reason neglecting gauge loop effects
in the gap equation is a much better approximation at the GUT
scale than at the weak scale (assuming a comparable cutoff).

For masses $m = O(M_{\rm GUT})$ the two parts of the effective potential
eq.~\refe{v_eff} are obviously of very different order of magnitude~:
$V_{\rm gap} = O(\De^2 M_{\rm GUT}^2)$ and $V_{\rm D} = O(M_{\rm GUT}^4)$.
Therefore large VEVs can only be generated in directions with vanishing
D--term (flat directions). This is required anyway in order not to break
SUSY at high energy. Our strategy will be first to determine the
flat directions and then solve eq.~\refe{gap_appr} restricted to them.

Some general remarks can be made about the condensate masses.
High scalar masses of $O(M_{\rm GUT})$ will only be induced
by the D--term contribution to the effective potential. The corresponding
massive fermions arise because of the fermion gaugino masses.
Since our D-term does not break supersymmetry, we may expect that the
heavy fields are those which become part of the massive
gauge multiplets of broken generators. The remaining light scalars
can receive masses of $O(\Delta^2)$ from the gap potential.
Masses of the same order for the fermionic partners are
implied by the tree level term in eq.~\refe{aux_lag}.

\section{Condensates in ${\bf r} \times \bf{\bar r}$}

In this section we consider the GUT breaking by adjoint condensates as is
favored for the usual GUT breaking. To get adjoint condensates we choose a
representation $\bf r$ and a complex conjugate representation $\bf \bar{r}$.
The composite fields from ${\bf r} \times {\bf \bar{r}}$ generally
consist of a singlet, an adjoint and possible other fields.
Tuning to a desirable condensate is now achieved by arranging the
coefficients $c_q$'s where $q$ runs for the singlet, the adjoint and
the other condensates.

First task is to pick up the flat directions.  When $c_q$ is taken to be as
large as to produce VEVs of $O(M_{\rm GUT})$, the absolute minimum of the
effective potential is sitting on the flat direction as can be seen from
eq.~\refe{pot_min}.  The obvious flat directions are
those with zero weights. Taking these, the mass matrix becomes
diagonal
\be \label{diagM}
   M_{ij} = \sum_{A, q} c_q m^{(q)A} d_{Ai}^{(q)} \delta_{ij}\; .
\ee
Here $d_{Ai}^{(q)}$ are the entries of the diagonal matrices $\la$
for a representation $q$.  Note that a field in the adjoint
representation can take the form of a diagonal matrix without
loss of generality. Therefore, if only a singlet and an adjoint are present
all flat directions are taken into account.
If there appear more representations the directions
with zero weights  do not cover all flat directions. Including the missing
directions makes the mass matrix off-diagonal. Since this complicates
the gap equation much we restrict the discussion to the zero weight
directions for simplicity. Since the mass matrix is now diagonal
one can find explicit solutions for the gap equation.
Let us now denote $M_{ij} = u_i \delta_{ij}$ with $u_i = c_q
m^{(q)A} d_{Ai}^{(q)}$.  Then the gap equation \refe{gap_appr} reads\footnote{
Here we put $\Lambda = 1$ and neglect the SUSY breaking parameter $\Delta$.}
\be \label{diagGapEq}
  u_i + \sum_j K_{ij} u_j (\ln |u_j|^2 +1) = 0, \qquad
  K_{ij} = \sum_q c_q^2 K^{(q)}_{ij}
\ee
where $K^{(q)}_{ij} = \sum_A d_{Ai}^{(q)} d_{Aj}^{(q)}$.  The matrices denoted
by $K^{(q)}$ are the projectors for the subspaces corresponding to the
representations $q$. They fulfill the following orthogonality
and completeness properties~:
\label{proj} \bea
  K^{(q)} K^{(p)} &=& \delta^{qp} K^{(p)} \\
  \sum_q K^{(q)}  &=& 1 \;.
\eea{gap_eq_ad}
To find the global minimum it is useful to notice the
existence of a bound of the gap potential.  Consider the case where only one
$c_q$ is non-zero and define $u_i = z_i \exp[-{1\over2}({1/c_q^2} + 1)]$.
The variables $z_i$ being in the space defined by $q$ should satisfy
the  constraint and  the gap equation  which read
\be
 \label{zGapCon}
  \begin{array}{rcl}
  \sum_j K_{ij}^{(q)} z_j &=& z_i\\
  c_q^2 \sum_j K_{ij}^{(q)} z_j \ln |z_j|^2 &=& 0 \;.
\end{array}
\ee
Multiplying $z_i$ on the both sides of the second equation results in
$\sum_i |z_i|^2 \ln |z_i|^2 = 0$.  Now one obtains the bound: $-V_{\rm gap}
\sim \sum_i |z_i|^2 \leq N$.  The value $N$ is taken if and only if
$|z_i| = 1$ for all $i$.
For the proof let us consider a $(N-1)$-dimensional surface defined by the
equation $\sum a_i \ln a_i = 0$ where $a_i$ denotes $|z_i|^2$. On this surface
the maximum value of $\sum a_i$ is given by $N$ when $a_i = 1$ for all $i$.
This can be shown by minimizing the function with Lagrange multiplier
$\lambda$: $f(a_i) = \sum a_i + \lambda \sum a_i \ln a_i$.
Minimization shows
that $\lambda = -1$ and $a_i = 1$ for all $i$. There can be possible minima on
the boundary. However the boundary is given by $a_i=0$ for some $i$ and
therefore described by a similar equation with lower $N$. Since our
statement holds for $N=2$ an induction in $N$ completes the proof.
Now let us go to the case with several $c_q$'s being nonzero.
First we normalize the $u_i$'s by taking the largest $c_q$,
say $c_l$: $u_i = z_i \exp[-{1\over2}({1\over c_l^2} +1)]$.  Then the gap
equation becomes
\be
 z_i + \sum K_{ij} z_j ( \ln|z_j|^2 - {1\over c_l^2}) = 0\;.
\ee
Multiplying the inverse K-matrix $K^{-1} = \sum_q {1\over c_q^2} K^{(q)}$ and
$z$ to the above equation leads to
\be
  \sum_i |z_i|^2 \ln |z_i|^2 = \sum_q ( {1\over c_l^2} - {1\over c_q^2})
  |z^{(q)}|^2 \label{bound}
\ee
where $z^{(q)} = K^{(q)}z$.
Since $c_l \ge c_q$ we have $\sum a_i \ln a_i \le
0$ with $a_i = |z_i|^2$. The absolute maximum value of $\sum a_i$ under
this condition is again given by $N$ and it is taken if and only if
$a_i=1$ for all $i$. Comparing with eq.~\refe{bound} shows that the
existence of such a solution implies the existence of a $z$ with
$|z_i|=1$ for all $i$ and $z = K^{(l)} z$. Obviously also the converse
holds. If one, therefore, finds a solution to these two constraints
it represents the absolute minimum of the gap potential.
These minima correspond to condensate formation in the representation
with the largest $c_q$. If some $c_q$'s take the same
value the above argumentation holds for the sum of the
corresponding representations.

Clearly, for the singlet part such a solution always exists since its
$K$ matrix is given by
\be
 K_{ij}^{\rm sin} = 1/d
\ee
where $d={\rm dim}({\bf r})$. This implies that for any representation
${\bf r}$ a dominating singlet coupling leads to a vacuum pointing
into the singlet direction.
Indeed, we find that it also exists for the other condensates formed
out of the fundamental representations of $SU(N)$ and $SO(N)$.

\underline{{\bf N} of $SU(N)$}:  Here we can have two condensates: singlet and
adjoint. The adjoint projection matrix reads
\be
 K_{ij}^{\rm adj}=\de_{ij}-K_{ij}^{\rm sin}=\de_{ij}-1/d \; .
 \label{k_sun}
\ee
Depending on the couplings for the two condensates, denoted by
$c_{\rm sin} $ and $c_{\rm adj}$, we find three different patterns of
solutions. If $c_{\rm sin} $ dominates only a singlet solution
\be
 u_i = \exp[-{1\over2}(1/c_{\rm sin}^2 + 1)] \quad \mbox{for all}\; i
\ee
is allowed. If $c_{\rm adj}$ is larger, the solution
takes the form
\be
 u_i = z_i \exp[-{1\over2}(1/c_{adj}^2 + 1)],\quad {\rm with}\;
 \sum_i z_i = 0 \, .
\ee
where the $z_i$'s have modulus 1.
Aside from a global phase, we have $N-2$ degrees of freedom which give rise to
degenerate vacua.  Depending on the number of equal $z_i$ $SU(N)$ can be
broken down to $\Pi_p (S)U(n_p)\times U(1)^s$ where $\sum_p n_p =N$
and the $n_p$ are constrained by $n_{\rm max} \leq \sum_{\rm others} n_p$.
The number $s$ is chosen to complete the rank of the group to $N-1$.
For instance, $SU(5)$ can be broken down to $U(1)^4$, $SU(2) \times U(1)^3$
or $SU(2)^2\times U(1)$.
Finally in the case of $c_{\rm sin} = c_{\rm adj}$ each phase $z_i$
can take an arbitrary value. Here, the breakdown of $SU(5)$ to SM can be
possible on the contrary to the previous case.

\underline{{\bf N} of $SO(N)$}:  We restrict ourselves to even values
of $N$~\footnote{For $SO(N)$ with $N$ odd the breaking patterns are
similar to $SO(N-1)$ apart from an additional $U(1)$ factor.}.
The condensates of ${\bf N}\times{\bf N}'$
consist of a singlet, an anti-symmetric, and a symmetric part. The projection
matrices are given by
\bes
 K^{\rm anti-sym}&=&{\rm Diag}\left( \left( \ba{cc} 1&1\\1&1 \ea \right),
               \cdots,\left( \ba{cc} 1&1\\1&1 \ea \right) \right) \nn \\
 K^{\rm sym}&=&{\rm Diag}\left( \left( \ba{cc} 1&-1\\-1&1 \ea \right),
               \cdots,\left( \ba{cc} 1&-1\\-1&1 \ea \right) \right) \; .
\ees
If $c_{\rm sin}$ is the largest one finds a singlet solution as above.
For dominating $c_{\rm anti-sym}$ the solution takes the form
\be
 z = (e^{i\varphi_1}, \cdots , e^{i\varphi_{N/2}}, -e^{i\varphi_1}, \cdots
      , -e^{i\varphi_{N/2}})
\ee
where the values of $\varphi$'s are arbitrary since the traceless condition
is already fulfilled.  The largest unbroken subgroup is
$SU({N\over2}) \times U(1)$ if all phases are the same.
Depending now on the choice of $\varphi$'s $SU({N\over2})$
can be broken further without any restriction.
If $c_{\rm sym}$ exceeds the other couplings, the solutions are given by
\be
 z = (e^{i\varphi_1}, \cdots , e^{i\varphi_{N/2}}, e^{i\varphi_1}, \cdots
      , e^{i\varphi_{N/2}}), \quad {\rm with}\;  \sum_p e^{i\varphi_p} = 0\; .
\ee
They result in the breaking patterns
$ SO(N) \rightarrow \Pi_p SO(2n_p)$ where $\sum_p n_p = {N \over 2}$
with $n_{\rm max} \leq \sum_{\rm others} n_p$.

For the $SO(10)$ case and a dominating $c_{\rm anti-sym}$
the interesting patterns
\[
  SO(10) \rightarrow \left\{ \ba{l} SU(5) \times U(1)'\\
                       SU(3) \times SU(2) \times U(1) \times U(1)' \ea \right.
\]
are allowed.
Here we interpret $\bf 10$ and $\bf 10'$ as exotic fields to cause a breaking
of $SO(10)$.  In the next section, we will consider the condensates ${\bf 10}
\times {\bf 16}$ and ${\bf 16} \times {\bf 16}$ for a more realistic
model with standard fermions and a broken $U(1)'$.
Observe that for condensates in the symmetric
representation $\bf 54$ the conventionally allowed Pati-Salam path
$$ \begin{array}{lcl}
  SO(10) &\rightarrow& SO(6) \times SO(4)\\
             &\sim&    SU(4) \times SU(2) \times SU(2)
 \end{array}
$$
is excluded.\\

So far all our examples had a highly degenerate vacuum with almost all of
its points leading to the unbroken subgroup $U(1)^r$ ($r$ is the rank
of the group). Only for specific choices larger unbroken groups could be
obtained. This behavior is related to the existence of our ``general''
solution discussed above. As we will see for higher representations
${\bf r}$ it does not exist any longer in general and the vacuum structure
changes. We restrict ourselves to the case where ${\bf r}$ is a second
rank tensor of $SU(N)$ in
\[
 {\bf N}\times {\bf N} = \left( \frac{(N-1)N}{2}\right)_A
                         +\left( \frac{(N+1)N}{2}\right)_S\, .
\]
Especially this may be of interest for the ${\bf 10}_A$ representation
in the standard $SU(5)$ GUT. Moreover we consider condensates in the
singlet or adjoint part of ${\bf r}\times {\bf \bar{r}}$ only.
Couplings $c_q$ of higher representations are set to zero. As a
first step it is useful to work out the constraints on the vacuum
which follow from this requirement. Here we denote the diagonal entries
of the mass matrix by $w_{ij}^{(S)}$, $i\ge j$ and $w_{ij}^{(A)}$, $i>j$ for
the symmetric and antisymmetric part, respectively. The corresponding
matrices $d$ can be expressed by those of the fundamental representations
like $d_{a,(ij)}^{(S)}\sim d_{ai}+d_{aj}$, $i\ge j$ and
$d_{a,(ij)}^{(A)}\sim d_{ai}+d_{aj}$, $i>j$. This can be used to compute
the projection matrices $K$~:
\bes
 K_{(ij),(kl)}^{(A)}&=&\frac{1}{N-2}(K_{ik}+K_{il}+K_{jk}+K_{jl})\quad
                      i>j,\; k>l \nn \\
 K_{(ij),(kl)}^{(S)}&=&\frac{1}{N+2}(K_{ik}+K_{il}+K_{jk}+K_{jl})\quad
                      i\ge j,\; k\ge l \; .
\ees
Here $K$ is given by eq.~\refe{k_sun}. Working out the constraints
$(K^{\rm sin}+K^{(A,S)})w^{(A,S)}=w^{(A,S)}$ leads to the
parameterization
\bes
 w_{ij}^{(A)}&=&\frac{1}{2}(u_i + u_j)\quad i>j \nn \\
 w_{ij}^{(S)}&=&\frac{1}{2}(u_i + u_j)\quad i\ge j
\ees
with arbitrary $u_i$. The breaking pattern can now directly read off
from the vector $u$. E.~g. the singlet part vanishes for $\sum_i u_i = 0$.
With this parameterization it can be shown that apart from trivial cases
(${\bf \bar{3}}\in {\bf 3}\times {\bf 3}$ of $SU(3)$
and ${\bf 6}\in {\bf 4}\times {\bf 4}$ of $SU(4)\sim SO(6)$) our standard
solution $|w_{ij}|=1$ for all $i,j$ does not exist for the adjoint.
The gap equation and the gap potential are given by
\be
 \sum_j (\zeta_i +\zeta_j)\ln\frac{|\zeta_i +\zeta_j|^2}{4}
  +2\ep\zeta_i \ln|\zeta_i|^2 = -\frac{2(N+\ep)}{N}\bt\sum_j \zeta_j
\ee
and
\be
 V\sim \sum_{(ij)}|\zeta_i + \zeta_j |^2 \left( \ln\frac{|\zeta_i
      +\zeta_j|^2}{4} - 1\right) +\frac{2(N+\ep)}{N}\bt |\sum_j \zeta_j |^2
 \label{sunn_pot}
\ee
with the rescaled quantities $u_i=\zeta_i \exp(-1/2(1/c_{adj}^2-1))$ and
$\ep = +1$ or $-1$ for the symmetric or antisymmetric part, respectively.
Both equations only depend on one combination of the couplings, namely
\[
 \bt=\frac{1}{c_{sin}^2}-\frac{1}{c_{adj}^2}\; .
\]
Let us first look at the case $c_{\rm sin}=0$. The potential then consists
of the first term in eq.~\refe{sunn_pot}. A numerical minimization
for small values of $N$ results in the following pattern for the
absolute minimum
\be
 \ba{llllll}
  \zeta_i&=&\rho_1&{\rm for}&n_1&\zeta_i \\
  \zeta_i&=&-\rho_2&{\rm for}&n_2&\zeta_i \\
  \zeta_i&=&\rho_p e^{i\varphi_p}&{\rm for}&n_p&\zeta_i \\
  \zeta_i&=&\rho_p e^{-i\varphi_p}&{\rm for}&\bar{n}_p&\zeta_i \ea
\ee
where $p=3,4$, $\rho_p\in {\bf R}^+$ and $\varphi_p\ne 0,\pi$.
This causes a symmetry breaking
\be
 SU(N)\ra SU(n_1)\times SU(n_2)\times \Pi_p SU(n_p)\times SU(\bar{n}_p)\; .
\ee
The numbers $n_p$ are given in table~\ref{numbers}.
\begin{table}
 \begin{center}
  \begin{tabular}{|c|c|c|}
   \hline
    &antisymm.&symm. \\
    $N$&$(n_1,n_2,n_3=\bar{n}_3)$&$(n_1,n_2,n_3,\bar{n}_3,n_4,\bar{n}_4)$ \\
    \hline \hline
    $3$&$(1,0,1)$&$(1,0,1,1,0,0)$ \\ \hline
    $4$&see $SO(6)$&$(2,0,1,1,0,0)$ \\ \hline
    $5$&$(3,0,1)$&$(1,0,1,1,1,1)$ \\ \hline
    $6$&$(4,0,1)$&$(2,0,2,2,0,0)$ \\ \hline
    $7$&$(5,0,1)$&$(3,0,2,2,0,0)$ \\ \hline
    $8$&$(5,1,1)$&$(3,0,3,0,2,0)$ \\
    &$(4,0,2)^*$& \\ \hline
   \end{tabular}\\[0.5cm]
  \end{center}
 \hskip 3.7 cm $^*${\footnotesize The two patterns are degenerate.}
  \caption{Vacuum patterns for adjoint condensates formed out of
           symmetric or antisymmetric second rank $SU(N)$ tensors.}
  \label{numbers}
\end{table}
For the ${\bf 10}_A$ of $SU(5)$ e.~g. the induced breaking is
$SU(5)\ra SU(3)\times U(1)^2$. We stress that -- contrary to the
case of fundamental representations -- we arrive at unique minima.

As an example of what can happen for $c_{\rm sin}\ne 0$ we take
again ${\bf 10}_A$ of $SU(5)$. If $\bt<0$ (the singlet coupling
dominates) we get a pure singlet solution with $\zeta_i=\exp(-\bt /2)$.
In the opposite case the pattern of $\zeta_i$ and the unbroken group
are unchanged with respect to the situation for $c_{\rm sin}=0$.
However, there appears a small singlet admixture which vanishes
for $\bt\ra\infty$ ($c_{\rm sin}\ra 0$). This is shown in fig.~\ref{mix}.
\begin{figure}
\setlength{\unitlength}{1cm}
\begin{picture}(1,0.5)(-1.0,-3.0)
\put(0,0){$|u_{\rm sin}|/|u|$}
\end{picture}
\begin{picture}(0.5,0.5)(-7.0,0.2)
\put(0,0){$\bt$}
\end{picture}
\caption{Dependence of the singlet strength $|u_{\rm sin}|/|u|$ on $\bt$.}
\label{mix}
\end{figure}
To conclude this section we remark that in all our examples the
fundamental fields receive high masses $\sum_i u_i \phi_i' \phi_i$.
Therefore, e.~g. the weak doublets in a ${\bf 10}$ of $SO(10)$ used
to form condensates cannot serve as low energy Higgs fields.
Composite fields as Higgs fields will be discussed in section 5.

\section{Condensates in ${\bf r}\times {\bf r}$ -- the examples $SO(10)$,
         $E_6$}

Unfortunately the flat directions for complex representations
cannot be isolated in general. Moreover, since the mass matrix $M$ will
not be diagonal any longer the consideration of all flat directions
needs the calculation of logarithms of large matrices
(e.~g.~$27\times 27$--matrices in the $E_6$--case).
Therefore we restrict the discussion to the most important
examples $SO(10)$ and $E_6$. Our method will be more pragmatic and
guided by phenomenology~: Only the flat directions among the SM-singlets
will be taken into account.

Unlike to the above section with ${\bf r} \times \bar{\bf r}$ condensates, we
have here different features of the flat directions and the solutions of
the gap equation.  The vacuum points into a D--flat direction if
contributions from two or more irreducible representations are arranged
to cancel each other. As a consequence we may expect some restrictions
on the couplings $c$.
Let us, as a toy example, consider the case with condensates from
${\bf 3} \times {\bf 3}$ of $SU(3)$.  We introduce two couplings
$c_{\bar{\bf 3}}$ and $c_{\bf 6}$ for $\bar{\bf3}$ and $\bf 6$
respectively.  Then, one can show that the flat direction is
given by vacuum expectation values for the condensates
$\bar{\bf 3}$ and $\bf 6$ satisfying $v_{{\bf 6},ij} =
\de^k_i \de^k_j v_{\bf \bar{3}}^k/\sqrt{2}$~\cite{buccella}.
The equality should also be compatible with the gap equation to produce the
high scale. In general, this can be obtained in some restricted range of the
parameter space for $c_{\bar{\bf3}}$ and $c_{\bf 6}$ only.

Since the $SO(10)$ model has to include fermions in the ${\bf 16}$
representation we consider ${\bf 16}\times {\bf 16}$ condensates.
However, we will see that as the only condensates they possess no
flat SM direction. The most economical way out is to add particles
in the ${\bf 10}$ which leads to condensates in the
${\bf 16}\times {\bf 10}$ and ${\bf 10}\times {\bf 10}$. The latter were
already discussed in section 3.
$E_6$ is known as the unique GUT group which admits a spontaneous
breakdown with mass giving fields only~\cite{bn}.
For that reason we will
-- in contrast to our other examples -- not discuss adjoint breaking of $E_6$
but concentrate ourselves on condensates in ${\bf 27}\times {\bf 27}$.\\

We begin our analysis with $E_6$ and take ${\bf r} = {\bf 27}$.
Condensates can be formed in the representations
\[
 {\bf 27}\times {\bf 27} \rightarrow ({\bf \bar{27}} + {\bf \bar{351}})_S
   + {\bf \bar{351}}_A .
\]
Since we consider one family only the two symmetric parts
appear. Then the Lagrangian possesses the structure
\be
 {\cal L} = \int d^2 \theta d^2 \bar{\theta} \left[
   {\bf \bar{27}}\,{\bf 27} + \frac{c_H^2}{f^2}|({\bf 27}\times {\bf 27})_
   {\bf \bar{27}}|^2 + \frac{c_\Phi^2}{f^2}|({\bf 27}\times {\bf 27})_
   {\bf \bar{351}}|^2 \right] . \label{e6_lag}
\ee
For the explicit calculation we use the maximal subgroup
$SU(3)_L\times SU(3)_R\times SU(3)_C$
of $E_6$. Notation and group properties are explained in the appendix.

A Lagrangian of the above type but with specific values for $c_H^2$ and
$c_\Phi^2$ is obtained in a coset model based on $E_7/E_6\times U(1)$.
General expressions for the K\"ahler potential of coset models
involving $E_6$ can be found
in the literature~\cite{ach_delduc}. The scaleless version
produces no ${\bf 351}$ part. For unbroken $U(1)$ charge
expansion up to quartic terms and rearranging the expression
according to the representation contents gives eq.~\refe{e6_lag} with
\[
 c_H^2 = 4\quad\quad c_\Phi^2 = -1 .
\]
Since $c_\Phi^2$ is negative no ${\bf 351}$ condensates are formed.
But with VEVs in the ${\bf 27}$ only no SM invariant flat directions
occur (see below). Therefore a breakdown at high energy is impossible.

In the form of eq.~\refe{aux_lag} the Lagrangian looks like
\bes
 \cal{L} &=& \int d^2 \theta d^2 \bar{\theta} \left[ \left( \bar{L}_\al^a
             L^\al_a\; + \;2\, {\rm perm.} \right) + \left(
             \bar{H}^{'\al}_a H^{'a}_\al\; + \; 2\,
             {\rm perm.} \right) \right. \nn \\
          && \left. + \left(\bar{\Phi}^{'\al}_a \Phi^{'a}_\al + \frac{1}{4}
             \bar{\Phi}^{'ab}_{\al\bt}\Phi^{'\al\bt}_{ab}+
             \bar{\Phi}^{'\al i}_{aj} \Phi^{'aj}_{\al i}\; + \; 2 \,
             {\rm perm.} \right) \right] \nn \\
          && + \left[ \int d^2 \theta \left[ H^{\al}_a \left(H^{'a}_\al -
             \frac{c_H}{\sqrt{5}f} \left( Q^i_\al \hat{Q}^a_i - \frac{1}{2}
             \ep^{abc}\ep_{\al\bt\ga}L^{\bt}_b L^\ga_c
              \right)\right) \right. \right. \nn \\
          &&  + \Phi^{\al}_a \left(\Phi^{'a}_\al -
             \frac{c_\Phi}{f}\sqrt{\frac{5}{6}} \left( \frac{1}{3}Q^i_\al
             \hat{Q}^a_i + \frac{1}{4}\ep^{abc}\ep_{\al\bt\ga}
             L^{\bt}_b L^\ga_c \right)\right) \label{e6lag1} \\
          && + \frac{1}{4}\Phi_{\al\bt}^{ab} \left( \Phi^{'\al\bt}_{ab} -
             \frac{c_\Phi}{f}\frac{1}{\sqrt{2}} \left( L^\al_a L^\bt_b +
             L^\al_b L^\bt_a \right)\right) \nn \\
          && + \left. \Phi^{\al i}_{a j} \left(\Phi_{\al i}^{'aj} -
             \frac{c_\Phi}{f}\left( Q_\al^i \hat{Q}_j^a - \frac{1}{3}
             \de^i_j Q^k_\al \hat{Q}_k^a \right)\right)\; + \; 2 \,
             {\rm perm.} \right]  \nn \\
          && +\left. \; {\rm h.\,c.} \right] \; . \nn
\ees
Leptons, left and right handed quarks are named by $L$, $Q$ and $\hat{Q}$,
respectively while the composites are called ${\bf 27}\sim H$,
 ${\bf 351}_S \sim \Phi$. The two omitted parts of the Lagrangian
are obtained by the index permutation $\al\ra a\ra i\ra\al$.

We observe that the Lagrangian~\refe{e6lag1} possesses an anomalous
global $U(1)$--symmetry with charges $Q_{\rm gl}({\bf 27},H,\Phi ) =
(1,-2,-2)$. Obviously VEVs in $H$ or $\Phi$ break this U(1)
resulting in a large axion scale $f_a = O(M_{\rm GUT})$ which is
excluded by cosmological considerations. One way out may be that
higher dimensional operators with dimension less than 10 can break
the $U(1)$ symmetry. Even if they are strongly suppressed they can affect
the upper cosmological bound on the axion scale~\cite{holman_barr}.
The price will be that such an approximate $U(1)$ will not solve
the strong CP--problem any longer.

Seven SM invariant VEVs are contained in the two condensate
representations~:
\[
 \ba{llll} m_1 = H^{\al =3}_{a=3}&m_2 = H^{\al =3}_{a=2}&
           m_3 = \Phi^{\al =3}_{a=3}&m_4 = \Phi^{\al =3}_{a=2} \\
           m_5 = \Phi_{\al,\bt =3,3}^{a,b=3,3}&m_6 = \Phi_{\al,\bt =3,3}^
           {a,b=2,3}&m_7 = \Phi_{\al,\bt =3,3}^{a,b=2,2}& \ea
\]
Using a $SU(3)_R$
transformation which acts on the indices $a=2,3$ the VEV $m_6$ can be
rotated to zero. We sum up the six remaining VEVs multiplied with
the corresponding matrices $\la$ from the appendix and arrive at
\be
 |M|^2 = {\rm Diag}(A,A,B,C,C,C,D,D,D)
\ee
with
\begin{eqnarray*}
 A&=&\left( \ba{ccc} |m_{1-}|^2 + |m_{2-}|^2&0&0\\
                     0&|m_{1-}|^2&-m_{1-} m_{2-}^*\\
                     0&-m_{2-}m_{1-}^*&|m_{2-}|^2 \ea \right) \\
 B&=&{\rm diag}(0,|\tilde{m}_7|^2,|\tilde{m}_5|^2) \\
 C&=&{\rm diag}(0,0,|m_{1+}|^2+|m_{2+}|^2) \\
 D&=&\left( \ba{ccc} 0&0&0\\
                    0&|m_{2+}|^2&m_{2+} m_{1+}^*\\
                    0&m_{1+}m_{2+}^*&|m_{1+}|^2 \ea \right)
\end{eqnarray*}
and the definitions
\be
 \ba{lllp{1cm}lll}
  m_{1,2+}&=&\frac{1}{\sqrt{5}}(c_H m_{1,2} +
              c_\Phi \sqrt{\frac{2}{3}}m_{3,4})&
  &m_{1,2-}&=&\frac{1}{\sqrt{5}}(-c_H m_{1,2} + c_\Phi \sqrt{\frac{3}{2}}
  m_{3,4})\\
  \tilde{m}_{5,7}&=&\sqrt{2}c_\Phi m_{5,7}\; .&&&& \ea
  \label{mm_def}
\ee
The potential value at the minimum is given by
\[
 V_{\rm gap,min} = -f^2 (|\tilde{m}_5|^2 + |\tilde{m}_7|^2 +
                   4(|m_{1-}|^2+|m_{2-}|^2)+6(|m_{1+}|^2+|m_{2+}|^2)).
\]
Fortunately $|M|^2$ contains at most two by two blocks so that the
computation of the logarithm is possible. After some algebra one
obtains explicit gap equations for $m_1,..,m_7$. Since $m_5$ and $m_7$
decouple they can be chosen to their trivial value zero or to
\be
 \frac{|\tilde{m}_{5,7}|^2}{\La^2} = \exp (-1/c_\Phi^2 -1)
 \label{m5_gap_eq}
\ee
independently. The remaining VEVs are determined by
\be
 \left( \ba{c} m_{1,2+}\\m_{1,2-} \ea \right) = \frac{1}{5}
 \left( \ba{cc} (3 c_H^2 + 2 c_\Phi^2)l_+&2(c_\Phi^2 - c_H^2)l_-\\
                3(c_\Phi^2 - c_H^2)l_+&(3 c_\Phi^2 + 2 c_H^2)l_- \ea \right)
 \left( \ba{c} m_{1,2+}\\m_{1,2-} \ea \right)
 \label{mpm_gap_eq}
\ee
where
\[
 l_\pm = \ln\left(\frac{\La^2}{\De^2 + |m_{1\pm}|^2 + |m_{2\pm}|^2} \right) .
\]
These equations have to be combined with the flatness conditions which
we extract from the D--term expressions given in the appendix~:
\bes
 |H^3_3|^2 + |\Phi^3_3|^2 &=& 2 |\Phi^{33}_{33}|^2\nn \\
 |H^3_2|^2 + |\Phi^3_2|^2 &=& 2 |\Phi^{22}_{33}|^2 \label{e6_flatness}\\
 \bar{\Phi}^3_3 \Phi^3_2 + \bar{H}^3_3 H^3_2 &=&0 \nn .
\ees
As expected we obtain four (real) conditions corresponding to the four
broken $E_6$ generators which are SM singlets~\cite{buccella}. The
first two equations belong to the broken central charges. They can
be computed much easier using the charge values given in ref.~\cite{slansky}.

The VEVs - according to their symmetry properties - can be divided into two
groups~: the VEVs with (3,3) indices and the one with (3,2) indices.
These groups also show up in the structure of the above set of equations.
By direct calculation it can now be seen that gap equations and flatness
conditions are non compatible for nonzero VEVs in both groups.
Therefore we end up with two possible patterns~:
\begin{eqnarray*}
  1)&&\Phi^{33}_{33},\Phi^3_3,H^3_3 \ne 0,\;{\rm others\; zero}\\
  2)&&\Phi^{22}_{33},\Phi^3_2,H^3_2 \ne 0,\;{\rm others\; zero} .
\end{eqnarray*}
The two cases are completely symmetric with respect to the remaining
nontrivial equations~: One of the first two flatness conditions has
to be combined with one part of eq.~\refe{mpm_gap_eq}
(the one with index 1 or 2) and with one of the eqs.~\refe{m5_gap_eq}.
Counting the degrees of freedom~\footnote{One phase say the one of $H^3_3$
can be rotated away by gauge freedom. Then $\Phi^3_3$ has to be real
in order to fulfill the gap equation.} the system turns out to be
overdetermined by one equation. Consequently a relation between $c_H^2$
and $c_\Phi^2$ results which is shown in fig.~\ref{ch_cp_rel}.
\begin{figure}
\setlength{\unitlength}{1cm}
\begin{picture}(0.5,0.5)(-2.0,-3.0)
\put(0,0){$c_\Phi^2$}
\end{picture}
\begin{picture}(0.5,0.5)(-7.0,0.2)
\put(0,0){$c_H^2$}
\end{picture}
\caption{Relation between $c_H^2$ and $c_\Phi^2$ for $E_6$ breakdown.}
\label{ch_cp_rel}
\end{figure}
A breakdown only takes place if $c_H^2$ and $c_\Phi^2$ are chosen
to lie on this curve. We emphasize that this constitutes a fine
tuning problem~: Already small deviations from the curve produce
a D-term $O(M_{\rm GUT}^4)$ which destroys the minimum at large VEVs.

If, nevertheless, we accept the fine tuning the breaking pattern reads
\[
 E_6\times U(1)_{gl} \ra
   SU(2)_L\times SU(2)_R\times SU(4)_C \; .
\]
Some additional $Z_N$ symmetries arise from a mixing between $U(1)_{gl}$
and internal $U(1)$'s. Their charges are generated by
\[
 Q_N=n Q_A + \al Q_{\rm gl}
\]
with $(N,n,\al)=(2,0,1);(3,1,1)$. $Q_A$
corresponds to the anomalous part of the global $U(1)$--symmetries in
the MSSM~\cite{ibanez_ross}. It is given by $Q_A=-1/4(Q^r+Q^t)$ where
$E_6\ra SO(10)\times U_t(1)\ra SU(5)\times U_t(1)\times U_r(1)$.
The detailed behavior of the VEVs is complicated but for reasonable
small values of $c_H$ they follow eq.~\refe{vev_beh} quit closely.
Large masses for all exotic particles in
${\bf 27}$ except for the right handed neutrino (RHN) are guaranteed
(see eq.~\refe{e6lag1}). \\

The above analysis may be used to look for the breakdown of $SO(10)$ as
well. The relevant decompositions for $SO(10)$ are
\begin{eqnarray*}
 {\bf 16}\times {\bf 16} &=& ({\bf 10} + {\bf 126})_S + {\bf 120}_A\\
 {\bf 16}\times {\bf 10} &=& {\bf \bar{16}} + {\bf \bar{144}} .
\end{eqnarray*}
Without the terms forming adjoint condensates which were discusses in
the last section the Lagrangian reads
\bes
 {\cal L} &=& \int d^2 \theta d^2 \bar{\theta} \left[
   {\bf \bar{16}}\,{\bf 16} + {\bf \bar{10}}\,{\bf 10} +
   \frac{\tilde{c}_H^{2}}{f^2}|({\bf 16}\times {\bf 16})_
   {\bf 10}|^2 + \frac{\tilde{c}_\Phi^{2}}{f^2}|({\bf 16}\times {\bf 16})_
   {\bf 126}|^2 \right. \nn \\
 &&\left. +\frac{c_H^2}{f^2}|({\bf 16}\times {\bf 10})_{\bf \bar{16}}|^2
   +\frac{c_\Phi^2}{f^2}|({\bf 16}\times {\bf 10})_{\bf \bar{144}}|^2
   \right] .
 \label{so10_lag}
\ees
The explicit expression can be obtained as a part of the $E_6$
Lagrangian~\refe{e6lag1}. We denote the condensates by
$H\sim {\bf 16}$, $\tilde{H}\sim {\bf 10}$,
$\Phi\sim {\bf 144}$ and $\tilde{\Phi}\sim {\bf \bar{126}}$.
Then the model possesses two
independent global $U(1)$ symmetries. They can be split into an anomalous
and non anomalous part~: $Q_{\rm gl,an}({\bf 16},{\bf 10},H,\tilde{H},
\Phi,\tilde{\Phi})=(1,1,-2,-2,-2,-2)$ and $Q_{\rm gl,nonan}
({\bf 16},{\bf 10},H,\tilde{H},\Phi,\tilde{\Phi})=(4,-5,1,-8,1,-8)$.

The Lagrangian~\refe{so10_lag} plus an additional
$|{\bf 10}\times {\bf 10}|^2$ term can be interpreted as part of the
$E_8/ SO(10)\times SU(3)\times U(1)$ coset model~\cite{itoh_kugo}. In this
model, however, the ${\bf 10}\times {\bf 10}$ coupling constant as well
as $c_H$ and $c_\Phi$ are negative so that no consistent breakdown
can occur. This situation changes in the scaleless version discussed
by Ellwanger~\cite{ell} where
$c_{{\bf 10}\times {\bf 10}}^2/\tilde{c}_\Phi^2/c_H^2 = 1/2/1$ and
$c_\Phi^2 = 0$ is obtained.

Three SM invariants appear in the
condensate representations which correspond to $m_2$, $m_4$ and $m_7$ of
the $E_6$ example. The combinations $m_\pm$ can be defined like in
eq.~\refe{mm_def}. Independently the calculation can easily be
performed using the maximal subgroup $SU(5)\times U(1)$.
According to the decompositions
\bes
 {\bf 10}&\ra& {\bf 5}(2) + {\bf \bar{5}}(-2)\nn \\
 {\bf 16}&\ra& {\bf 10}(-1) + {\bf \bar{5}}(3) +{\bf 1}(-5) \label{so10br}
\ees
we write
\[
 \ba{ccccc}
  {\bf 16}&\sim&\psi&\sim&(\psi_{ij},\psi^i,\psi_0)\\
  {\bf 10}&\sim&\phi&\sim&(\phi_i,\phi^i). \ea
\]
The three SM invariants $\tilde{\Phi}_0$, $H_0$ and $\Phi_0$ are then given by
\[
 \ba{lllll}
 m_2&\sim&H_0&\sim&\psi^i\phi_i\\
 m_4&\sim&\Phi_0&\sim&
  2\sum_{\al=1}^{3}\psi^\al\phi_\al - 3\sum_{a=4}^{5}\psi^a\phi_a. \\
 m_7&\sim&\tilde{\Phi}_0&\sim&\psi_0\psi_0 \ea
\]
Reading off the matrices $\la$
\[
 \ba{lllp{1cm}lll}
 \la_H&=&\frac{1}{\sqrt{5}} \left( \ba{cc} 0&0\\ {\bf 1}&0\\0&0 \ea \right)
         \ba{c} {\bf 10}\\{\bf \bar{5}}\\{\bf 1} \ea &&
 \la_\Phi&=&\frac{1}{\sqrt{30}} \left( \ba{cc} 0&0\\ D&0\\0&0 \ea \right)
         \ba{c} {\bf 10}\\{\bf \bar{5}}\\{\bf 1} \ea \\
 &&\ba{p{0.5cm}cc} &{\bf 5}&{\bf \bar{5}} \ea &&&&
   \ba{p{0.7cm}cc} &{\bf 5}&{\bf \bar{5}} \ea \ea
\]
\begin{eqnarray*}
 \la_{\tilde{\Phi}}&=&{\rm diag}
     (\underbrace{0,...,0}_{10},\underbrace{0,...,0}_{5},
                          \sqrt{2}) \\
 D&=&{\rm diag}(2,2,2,-3,-3)
\end{eqnarray*}
we arrive at
\begin{eqnarray*}
 |M_{16\times 16}|^2&=&{\rm diag}(0,...0,2\tilde{c}_\Phi^{2} m_7^2)\\
 |M_{16\times 10}|^2&=&{\rm diag}(m_+^2,m_+^2,m_+^2,m_-^2,m_-^2,0,0,0,0,0).
\end{eqnarray*}
For $m_7$ we get
\be
 \frac{m_7^2}{\La^2} = \frac{1}{2\tilde{c}_\Phi^{2}} \exp
 (-1/\tilde{c}_\Phi^{2})
\ee
and for $m_\pm$ the $E_6$ result~\refe{mpm_gap_eq}. The flatness
condition is given by the second eq.~\refe{e6_flatness}. This also
can be seen from the $U(1)$ charges in eq.~\refe{so10br}.
The only difference
to $E_6$ is the appearance of a third independent coupling $\tilde{c}_\Phi^2$
which determines $m_7$. Therefore the structure of solutions is
similar~: Large VEVs are only generated if the three couplings fulfill
one constraint. For $c_\Phi^2 = \tilde{c}_\Phi^2$ this constraint is just
represented by fig.~\ref{ch_cp_rel}. The breaking chain then reads
\[
 SO(10)\times U(1)_{\rm gl,an}\times U(1)_{\rm gl,nonan}\ra
   SU(3)_C\times SU(2)_L\times U(1)_Y\times U(1)_{\rm gl,rem}
\]
where the remnant $U(1)$ is given by $Q_{\rm gl,rem} =
3Q_R - Q_{\rm gl,an} + Q_{\rm gl,nonan}$ and $Q_R=1/5(Q^r+Y)$.
Moreover discrete symmetries remains which we did not
note explicitly. Observe that the additional gauged $U(1)'$ left
over in the case of adjoint condensates only is now broken.

Contrary to the $E_6$ example the RHN receives a heavy masses $m_7$
and consequently no exotic fundamental field remains light.\\

We finish this section with a remark about the fine tuning problem.
Suppose in the $E_6$ example an additional fermion--mirror fermion pair
${\bf 27'}$ and ${\bf\bar{27}'}$ is introduced. The condensates in
${\bf 27'}\times {\bf\bar{27}'}$ can generate heavy masses for this new
fields. All complex condensate formation in ${\bf 27}\times {\bf 27}$,
${\bf 27'}\times {\bf 27}'$ and ${\bf\bar{27}'}\times {\bf\bar{27}}'$
with couplings $(c_H,c_\Phi)$, $(c_H',c_\Phi')$ and
$(\bar{c}_H',\bar{c}_\Phi')$ is described by the equations~\refe{m5_gap_eq}
and~\refe{mpm_gap_eq}. Choosing now $(c_H,c_\Phi)=(\bar{c}_H',\bar{c}_\Phi')
\ne (c_H',c_\Phi')$ produces a vacuum in a D--flat direction since the
mirror contribution enters the D--term with the opposite sign. Moreover
all SM invariants can receive large VEVs allowing for a direct breaking
to the SM~\footnote{Here we assume that the breaking generated by condensates
in ${\bf 27'}\times {\bf\bar{27}'}$ leaves the SM invariant. This can e.~g.
be achieved for a dominating singlet coupling.}. An analogous modification
is possible for $SO(10)$. In principle, of course, setting
$(c_H,c_\Phi)=(\bar{c}_H',\bar{c}_\Phi')$ is a fine tuning as serious as
the former one. However, such a relation might be easier to realize
in terms of an underlying theory than the complicated one represented
in fig.~\ref{ch_cp_rel}.

\section{Masses of condensates and phenomenological discussion}

The general structure of condensate masses was already mentioned in the
end of section 2. Let us now be more specific in order to discuss the
possibility of radiative induced breakdown of the low energy theory
triggered by condensates.

Clearly the gauge singlets in ${\bf r}\times {\bf \bar{r}}$ receive no
large mass. For the adjoint part it is easy to show that any supermultiplet
which corresponds to a broken generator contains one (real) massive mode
and a goldstone boson. Looking at our examples we see that there exists only
one way to achieve a breakdown of $SU(5)$ to the SM~: A ``degeneracy'' in
the singlet and adjoint direction of ${\bf 5}\times {\bf \bar{5}}$
has to be implemented by setting $c_{\rm sin}=c_{\rm adj}$. Taking
into account gauge corrections, however, we expect this degeneracy
to be lifted in favor of the singlet direction. The smallest groups
which can lead to SM($\times U(1)'$) by ${\bf r}\times {\bf\bar{r}}$
condensates are therefore $SU(6)$ and $SO(10)$. But even if the
problem of a highly degenerated vacuum in these cases will be solved
by higher radiative corrections no candidates for the low energy
Higgses are present with adjoint condensates only.
A discussion of condensates contained in complex representations is
therefore in order.

 From the D-term expressions in the appendix we can read off the 57
massive condensate modes for the $E_6$ model~:
\[ \ba{l}
 m_1^* H^3_3+m_3^* \Phi^3_3-2m_5\bar{\Phi}^{33}_{33}\;+\;{\rm h.c.} \\
 m_1^* H^\al_3+m_3^*\Phi^\al_3-\sqrt{2}m_5\bar{\Phi}^{\al 3}_{33}\quad\quad
 m_1^* H_a^3+m_3^*\Phi_a^3-\sqrt{2}m_5\bar{\Phi}_{a3}^{33}\quad\quad
  \al ,a=1,2\\
 m_1^*H^b_i+\frac{2}{3}m_3^*\Phi^b_i-\frac{\sqrt{5}}{3}m_3^* \Phi^{b3}_{i3}
 +\sqrt{2}m_5\bar{\Phi}^{3b}_{i3}\quad\quad b=1,2\quad i=1,2,3 \\
 m_1^*H_\bt^i+\frac{2}{3}m_3^*\Phi_\bt^i-\frac{\sqrt{5}}{3}m_3^*
 \Phi_{\bt 3}^{i3} +\sqrt{2}m_5\bar{\Phi}_{3\bt}^{i3}\quad\quad \bt = 1,2\quad
 i=1,2,3 \\
 \ba{lll}
 m_3^* u+m_3 \bar{v}&{\rm with}&(u,v)=(\Phi^{ia}_{\ga 3},
  \Phi^{c3}_{i\al}) \\
  &{\rm and}&(a\ga,c\al)=(11,22);(22,11);(12,21);(21,12)\quad i=1,2,3\; .
  \ea \ea
\]
One uncolored goldstone mode is given by the imaginary part corresponding
to the first expression. The others are obtained from the
second and third expression by changing the sign of the last
term~\footnote{Their orthogonality to the massive modes is guaranteed
by the flatness condition.}.
We do not quote the colored goldstone modes which are a bit more complicated.
The above results hold for the case $m_1,m_3,m_5\ne 0$.
The other case is obtained by exchanging $a=3\leftrightarrow a=2$.
First we remark that the remaining
SM singlets $m_2,m_4,m_7$ stay massless. They may induce a further
breakdown to SM. If this happens at a scale $\sim 10^{11}$ GeV a RHN
mass desirable for the MSW solution of the solar neutrino problem
is generated. Obviously both pairs of weak doublets which couple to the
standard particles remain light. Clearly also in our modified $E_6$
model with mirror fermions light weak doublets are present.

The $SO(10)$ result for ${\bf 16}\times {\bf 16}$ and ${\bf 16}\times {\bf 10}$
corresponds to the above $E_6$ result for $m_2,m_4,m_7\ne 0$. A light
pair of weak doublets is guaranteed. We see that the $SO(10)$ model with
${\bf 16}$ and ${\bf 10}$ fields possesses the desired features. Since
the adjoint condensate in ${\bf 10}\times {\bf 10}'$ renders the ${\bf 10}$
superheavy all condensate formation has to take place at the GUT scale.
The RHN mass will therefore be of $O(M_{\rm GUT})$.

So far we did not discuss other condensate fields than weak doublets.
We emphasize that the number of chiral condensate multiplets which disappear
as massive or goldstone states equals the number of broken generators.
For realistic examples which need several condensate fields one may
therefore expect many new particles at the SUSY breaking scale.
Especially the color triplets belonging to the weak doublets
remain light.
This is no surprise because we could not expect the dynamical
mechanism to generate a doublet--triplet splitting which in ordinary
GUTs is imposed ``by hand''. Proton decay via triplet exchange needs
a mixture of triplet and antitriplet. Though our global remnant symmetry
forbids such a mixture for $SO(10)$ we expect it to be induced
after elektroweak breakdown. Even if this occurs in higher loop order
it will destabilize the proton at an unacceptable rate.

\section{Conclusion}

In this paper we analyzed the dynamical breakdown of supersymmetric
GUTs triggered by NJL condensates. Motivated by the standard GUT
breaking with adjoint representations we first looked at condensates in
${\bf r}\times {\bf \bar{r}}$. For ${\bf r}$ being the
fundamental representation of $SU(N)$ or $SO(N)$ we found general
results~: The vacuum is contained in the representation specified by
the largest coupling. It turns
out to be highly degenerated breaking to $U(1)^r$ ($r$ is the rank) for
almost all of its points. Choosing special points in the vacuum can
produce larger unbroken groups. For $SU(5)$, however, only
$SU(2)\times U(1)^3$ is admitted whereas $SO(10)$ can be broken to
SM$\times U(1)'$. The behaviour of condensates of higher rank
tensors we found to be very different. Adjoint condensates from second
rank $SU(N)$ tensors develop a unique vacuum and a small singlet admixture
occures for dominating adjoint coupling. The ${\bf 10}_A$ of $SU(5)$ e.~g.
generates a breakdown to $SU(3)\times U(1)^2$.

A realistic dynamical breaking cannot be achieved using adjoint breaking
only. First of all $SU(5)$ cannot be broken to the SM in this
way (using the small representations). Starting with larger groups,
however, forces one to break the additional $U(1)$ factors
and to give mass to the exotic fermions. Moreover, with an adjoint
only no candidates for the low energy Higgses are present.

Therefore we investigated complex condensates for the examples $SO(10)$
and $E_6$. Since we restricted the calculation to the SM invariant
directions in a strict sence our results only give necessary but not
sufficient conditions for the determined breakdown.
To guarantee D-term flatness in these cases VEVs in
different representations have to cancel each other. We only found nontrivial
solutions under this requirement if certain relations between the
couplings hold. As a consequence the couplings have to be
fine tuned to obtain a breakdown at the GUT scale. Accepting this
$E_6$ with the fermion field in $\bf 27$
can be broken to $SU(4)_C\times SU(2)_R\times SU(2)_L$.
A breakdown of $SO(10)$ to the SM may be achieved by taking
an exotic field in $\bf 10$ as well as the fermion field in $\bf 16$.
Typically our models possess
additional global symmetries which after the breakdown lead to a
large scale axion. Combinations of these symmetries with gauged
$U(1)$ factors can remain in the low energy theory. For $SO(10)$ e.~g.
we find a remnant $U(1)$.
Alternatively we discussed a model with additional mirror fermions.
Then setting coupling constants and mirror coupling constants equal guarantees
D--term flatness. In this scenario $E_6$ can be broken to the SM directly.
In all cases we find light composite doublets which
can trigger a further breaking.

Large composite masses are generated through gauge terms only.
Therefore the number of composite chiral fields which disappear
from the low energy theory equals the number of broken generators.
Consequently a large number of composite particles at the SUSY breaking scale
can be expected in typical examples. Especially we find that the
color triplets accompanying the weak doublets remain light.
They lead to fast proton decay and are
clearly unacceptable. Therefore solving the doublet--triplet problem is
no longer merely an aesthetical problem but necessary to
obtain a realistic theory.\\[0.5cm]
{\bf Acknowledgement}~We are grateful to Y.~Achiman and H.~P. Nilles
for discussions and suggestions.  A part of this work was performed
when EJC was visiting the Physics Department of the National Technical
University in Athens. He thanks for its hospitality.\\[1.5cm]
\appendix{\large \bf Appendix}
 \renewcommand{\theequation}{\Alph{section}.\arabic{equation}}
\setcounter{equation}{0}
\section{$E_6$ group properties and notation}
In this appendix we summarize the main group properties of $E_6$
used in section 4 in terms of the subgroup $SU(3)_L\times SU(3)_R
\times SU(3)_C$. The $SU(3)$--factors are indexed by $\al,\bt,...$,
$a,b,...$ and $i,j,...$, respectively. Usually only one third
of the results is written explicitly. The other two parts can be obtained
by the permutation $\al\ra a\ra i\ra\al$.
The $E_6$ representations decompose like
\begin{eqnarray*}
 {\bf 27}&\ra& ({\bf\bar{3}},{\bf 3},{\bf 1})+({\bf 3},{\bf 1},{\bf\bar{3}})
               +({\bf 1},{\bf\bar{3}},{\bf 3})\\
 {\bf 78}&\ra& ({\bf 8},{\bf 1},{\bf 1})+({\bf 1},{\bf 8},{\bf 1})
               +({\bf 1},{\bf 1},{\bf 8})+({\bf 3},{\bf 3},{\bf 3})
               +({\bf\bar{3}},{\bf\bar{3}},{\bf\bar{3}})\\
 {\bf 351}_S&\ra&({\bf\bar{3}},{\bf 3},{\bf 1})+({\bf 6},{\bf\bar{6}},{\bf 1})
                 +({\bf\bar{3}},{\bf 3},{\bf 8})\; + \; 2\, {\rm perm.}
\end{eqnarray*}
We denote
\begin{eqnarray*}
 {\bf 27}&\sim& (u^\al_a,u^i_\al,u^a_i)\\
 {\bf 78}&\sim& (Q^\al_\bt,Q^a_b,Q^i_j,Q_{\al ai},Q^{\al ai})\\
 {\bf 351}_S&\sim& (w^\al_a,w^{ab}_{\al\bt},w^{\al i}_{aj},2\, {\rm perm.})\; .
\end{eqnarray*}
Using the $E_6$ algebra given in ref.~\cite{del_val} its action on the
${\bf 27}$ can be computed~:
\bes
 Q_{\al ai}(u^\bt_b)&=&-\de^\bt_\al \ep_{abc} u^c_i \nn \\
 Q^{\al ai}(u^\bt_b)&=&\de^a_b \ep^{\al\bt\ga} u^i_\ga
 \nn \\
 Q^\al_\bt (u^\ga_k)&=&-\de^\ga_\bt u^\al_c +\frac{1}{3}\de^\al_\bt
                      u^\ga_c \label{27_alg}\\
 Q^\al_\bt (u^k_\ga)&=&-\de^\al_\ga u_\bt^k -\frac{1}{3}\de^\al_\bt
                       u^k_\ga \nn\\
 Q^\al_\bt (u^c_k)&=&0\; .\nn
\ees
The various components of $u$ and $w$ can be expressed as bilinears
in $v\sim {\bf\bar{27}}$~\cite{schuecker}~\footnote{Observe that with
these definitions $w^{ab}_{\al\bt}$ and $w^{\al i}_{aj}$ are not
normalized to one for $a=b$, $\al=\bt$ or $i=j$. If we quote formulas
with specific values for the indices we reintroduce the proper
normalization.}:
\bes
 u^\al_a&=&\frac{1}{\sqrt{5}}\left( v^\al_i v^i_a - \frac{1}{2}
           \ep^{\al\bt\ga}\ep_{abc}v^b_\bt v^c_\ga \right)\nn \\
 w^\al_a&=&\sqrt{\frac{6}{5}}\left( v^\al_i v^i_a + \frac{1}{4}
           \ep^{\al\bt\ga}\ep_{abc}v^b_\bt v^c_\ga \right)\nn \\
 w^{ab}_{\al\bt}&=&\frac{1}{\sqrt{2}}( v^a_\al v^b_\bt+
                    v^b_\al v^a_\bt) \label{27_prod} \\
 w^{\al i}_{aj}&=&v^\al_j v^i_a - \frac{1}{3}\de^i_j v^\al_k v^k_a\; . \nn
\ees
 From these equations it is easy to read off the matrices $\lambda$
appearing in the Lagrangian~\refe{start_lag}~:
\begin{eqnarray*}
 \lambda_{\bf 27}(^a_\al)&=&\frac{1}{\sqrt{5}}\left( \ba{ccc}
                            -\ep^{abc}\ep_{\al\bt\ga}&0&0\\
                            0&0&\de^j_k\de^\ga_\al\de^a_b\\
                            0&\de^k_j\de^\bt_\al\de^a_c&0 \ea \right) \\
 \lambda_{\bf 351}(^a_\al)&=&\sqrt{\frac{6}{5}}\left( \ba{ccc}
                            \frac{1}{2}\ep^{abc}\ep_{\al\bt\ga}&0&0\\
                            0&0&\frac{1}{3}\de^j_k\de^\ga_\al\de^a_b\\
                            0&\frac{1}{3}\de^k_j\de^\bt_\al\de^a_c&0
                            \ea \right) \\
 \lambda_{\bf 351}(^{\al\bt}_{ab})&=&\frac{1}{\sqrt{2}}\left(\ba{ccc}
                  (\de^\al_\ga\de^\bt_\de+\de^\al_\de\de^\bt_\ga)
                  (\de^c_a\de^d_b+\de^c_b\de^d_a)&0&0\\
                  0&0&0\\
                  0&0&0 \ea \right) \\
 \lambda_{\bf 351}(^{al}_{\al i})&=&\left( \ba{ccc}
                  0&0&0\\
                  0&0&\de^\ga_\al\de^a_b (\de^l_k\de^j_i-\frac{1}{3}
                                          \de^l_i\de^j_k)\\
                  0&\de^\bt_\al\de^a_c (\de^l_j\de^k_i-\frac{1}{3}
                                          \de^l_i\de^j_k)&0 \ea \right)
\end{eqnarray*}
Moreover together with eq.~\refe{27_alg} they allow to compute the
action of the group generators on the ${\bf 351}$ components~:
\bes
 Q^\al_\bt (w^\ga_k)&=&-\de^\ga_\bt w^\al_c +\frac{1}{3}\de^\al_\bt
                      w^\ga_c \nn \\
 Q^\al_\bt (w^c_k)&=&0  \nn \\
 Q^\al_\bt (w^k_\ga)&=&\de^\al_\ga w_\bt^k -\frac{1}{3}\de^\al_\bt
                       w^k_\ga \nn \\
 Q^\al_\bt (w^{cd}_{\ga\de}) &=&\de^\al_\ga w^{cd}_{\bt\ga}+\de^\al_\de
           w^{cd}_{\ga\bt}-\frac{2}{3}\de^\al_\bt w^{cd}_{\ga\de} \nn \\
 Q^\al_\bt (w^{kl}_{cd})&=&0 \nn \\
 Q^\al_\bt (w^{\ga\de}_{kl})&=&-\de^\ga_\bt w^{\al\de}_{kl}-\de^\de_\bt
           w^{\ga\al}_{kl}+\frac{2}{3}\de^\al_\bt w^{\ga\de}_{kl} \nn \\
 Q^\al_\bt (w^{\ga\de}_{kl})&=&-\de^\ga_\bt w^{\al k}_{cl} +\frac{1}{3}
           \de^\al_\bt w^{\ga k}_{cl}  \label{351_alg} \\
 Q^\al_\bt (w^{c\ga}_{k\de})&=&\de^\al_\ga w^{c\ga}_{k\bt}-\de^\ga_\bt
           w^{c\al}_{k\de}\nn \\
 Q^\al_\bt (w^{kc}_{\ga d})&=&\de^\al_\ga w_{\bt d}^{kc} -\frac{1}{3}
           \de^\al_\bt w^{kc}_{\ga d} \nn\\
 Q_{\al ai}(w^\bt_b)&=&\frac{2}{3}\de^\bt_\al \ep_{abc}w^c_i-
                       \sqrt{\frac{5}{6}}\ep_{abc}w^{c\bt}_{i\al} \\
 Q_{\al ai}(w^{bc}_{\bt\ga})&=&-\frac{1}{\sqrt{2}}(\ep_{\al\bt\ga}
           (\de^b_a w^{c\de}_{i\ga}+\de^c_a w^{b\de}_{i\ga})+
           (\bt\lra\ga)) \nn \\
 Q_{\al ai}(w^{\bt j}_{bk})&=&-\sqrt{\frac{5}{6}}\de^\bt_\al \ep_{abc}
           (\de^j_i w^c_k-\frac{1}{3}\de^j_k w^c_i)-\frac{1}{\sqrt{2}}
           \de^\bt_\al \ep_{ikl} w^{jl}_{ab} \nn \\
           &&-\ep_{abc}(\de^j_i w^{c\bt}_{k\al}-\frac{1}{3}\de^j_k
             w^{c\bt}_{i\al}) \nn\; .
\ees
The action of the $Q^{\al ai}$ is given analogously to that of the
$Q_{\al ai}$ but with all coefficients replaced by their negatives.
Finally all D--terms can be determined with the given
formulas~:
\bes
 D_{\bf 27}(^\al_\bt)&=&\bar{u}^\al_k u^k_\bt - \bar{u}^c_\bt u^\al_c
           +\frac{1}{3}(\bar{u}^c_\ga u^\ga_c - \bar{u}^\ga_k u^k_\ga)
           \de^\al_\bt \nn\\
 D_{{\bf 27},(\al ai)}&=&-\ep_{abc}\bar{u}^b_\al u^c_i - \ep_{\al\bt\ga}
            \bar{u}^\bt_i u^\ga_a - \ep_{ijk} \bar{u}^j_a u^k_\al \nn \\
 D_{{\bf 27},(\al ai)}&=&\ep^{\al\bt\ga} \bar{u}^a_\bt u^i_\ga +
            \ep^{ijk}\bar{u}^\al_j u^a_k + \ep^{abc}\bar{u}^i_b u^\al_c
            \label{d_terms} \\
 D_{\bf 351}(^\al_\bt)&=&-\bar{w}^c_\bt w^\al_c + \bar{w}^\al_k w^k_\bt +
            \frac{1}{2}(\bar{w}^{\al\de}_{cd}w^{cd}_{\bt\de} -
            \bar{w}^{kl}_{\ga\bt} w^{\ga\al}_{kl}) \nn\\
            &&-\bar{w}^{cl}_{\bt k}w^{\al k}_{cl} + \bar{w}^{k\al}_{c\ga}
            w^{c\ga}_{k\bt} - \bar{w}^{k\de}_{c\bt}w^{c\al}_{k\de} +
            \bar{w}^{\al d}_{kc}w^{kc}_{\bt d} \nn \\
            &&+\frac{1}{3}\de^\al_\bt (\bar{w}^c_\ga w^\ga_c -
            \bar{w}^\ga_k w^k_\ga - \frac{1}{2}\bar{w}^{\ga\de}_{cd}
            w^{cd}_{\ga\de} + \frac{1}{2}\bar{w}^{kl}_{\ga\de} w^{\ga\de}_{kl}
            + \bar{w}^{cl}_{\ga k}w^{\ga k}_{cl} - \bar{w}^{\ga d}_{kc}
            w^{kc}_{\ga d}) \nn \\
 D_{{\bf 351},\al ai}&=&\frac{2}{3}\ep_{abc}\bar{w}^b_c w^c_i-
            \sqrt{\frac{5}{6}}\ep_{abc}(\bar{w}^b_\bt w^{c\bt}_{i\al}
            +\bar{w}^{bk}_{\al i} w^c_k) \nn \\
            &&-\frac{1}{\sqrt{2}}(\ep_{\al\bt\de}\bar{w}^{\bt\ga}_{ac}
            w^{c\de}_{i\ga}+\bar{w}^{j\ga}_{a\bt}w^{\bt\de}_{ij})
            -\ep_{abc}\bar{w}^{bk}_{\bt i}w^{c\bt}_{k\al}+
            {\rm 2\, perm.} \nn\; .
\ees
Again the expression for $D_{\bf 351}^{\al ai}$ can be obtained as an
analog of the one for $D_{{\bf 351},\al ai}$ changing the coefficients
to their negative values.

\end{document}